\documentclass[aps,prl,twocolumn,superscriptaddress]{revtex4}
\usepackage{graphicx,epsfig}

\def \ba {\begin{eqnarray}}
\def \ea {\end{eqnarray}}
\def \bc {\begin{cases}}
\def \ec {\end{cases}}
\def \be {\begin{equation}}
\def \ee {\end{equation}}
\def \barr {\begin{array}}
\def \earr {\end{array}}
\def \bi {\begin{itemize}}
\def \ei {\end{itemize}}

\def \ra {\rightarrow}
\def \da {\downarrow}
\def \ua {\uparrow}

\newcommand{\llangle}{\langle\!\langle}
\newcommand{\rrangle}{\rangle\!\rangle}

\begin{document}

\title{Dynamical mean-field theory of Hubbard-Holstein model
at half-filling: Zero temperature metal-insulator and
insulator-insulator transitions}
\author{Gun Sang Jeon}
\affiliation{Department of Physics, BK21 Physics Research
Division, and Institute for Basic Science Research, Sung Kyun Kwan
University, Suwon 440-746, Korea.}
\affiliation{The Pennsylvania
State University, Department of Physics, University Park, PA
16802.}
\author{Tae-Ho Park}
\affiliation{Department of Physics, BK21 Physics Research
Division, and Institute for Basic Science Research, Sung Kyun Kwan
University, Suwon 440-746, Korea.}
\author{Jung Hoon Han}
\affiliation{Department of Physics, BK21 Physics Research
Division, and Institute for Basic Science Research, Sung Kyun Kwan
University, Suwon 440-746, Korea.} \affiliation{CSCMR, Seoul National
University, Seoul 151-747, Korea.}
\author{Hyun C. Lee}
\affiliation{Department of Physics and Basic Science Research
Institute, Sogang University, Seoul 121-742, Korea}
\author{Han-Yong Choi}
\affiliation{Department of Physics, BK21 Physics Research
Division, and Institute for Basic Science Research, Sung Kyun Kwan
University, Suwon 440-746, Korea.} \affiliation{CNNC, Sung Kyun Kwan University,
Suwon 440-746, Korea.} \affiliation{Asia Pacific Center for
Theoretical Physics, Pohang 790-784, Korea.}

\date{\today}

\begin{abstract}
We study the Hubbard-Holstein model, which includes both the
electron-electron and electron-phonon interactions characterized
by $U$ and $g$, respectively, employing the dynamical mean-field
theory combined with Wilson's numerical renormalization group
technique. A zero temperature phase diagram of metal-insulator and
insulator-insulator transitions at half-filling is mapped out
which exhibits the interplay between $U$ and $g$. As $U$ ($g$) is
increased, a metal to Mott-Hubbard insulator (bipolaron insulator)
transition occurs, and the two insulating states are distinct and
can not be adiabatically connected. The nature of and transitions
between the three states are discussed.

\end{abstract}

%\pacs{PACS numbers:~74.20.Mn, 75.10.Jm, 74.25.-q}

\maketitle

%{\it Introduction.} --
The interaction-induced phenomena are a
fundamental problem because of the new physics that emerges. For
instance, the interaction-driven metal-insulator transition (MIT)
has been studied to a great extent for systems which involve the
electron-electron or electron-phonon interactions \cite{imada}. In
real condensed-matter systems, both interactions exist and it will
be important to understand the interplay between them.
%Such interplay will be
%important to the problems such as the effects of phonons in the
%strongly correlated electron systems and the effects of electron
%correlations in the electron-phonon coupled systems. They include
%the cuprates, organic superconductors, Kondo systems, fullerenes,
%and the transition metal oxides, among others.
The prototype model for describing such interplay is the
Hubbard-Holstein (HH) model, which includes the onsite
electron-electron and electron-phonon interactions characterized
by $U$ and $g$, respectively, as given by Eq.\ (\ref{hhm}) below.
The Hubbard model ($g=0$) and Holstein model ($U=0$), which are
special limits of the more general HH model, have been extensively
studied using the dynamical mean-field theory (DMFT) in infinite
dimensions \cite{metzner,georges} in the context of the
interaction-driven MIT. Both models exhibit MIT as $U$ or $g$ is
increased above a critical value. Now, consider the MIT in the
$U-g$ plane. Natural questions then arise as to whether and how
the two insulating states, metallic states, and metal-insulator
transitions of the Hubbard and Holstein models are different, and
how they are affected when both $U$ and $g$ are present and
compete with each other. These questions will be addressed here by
studying the HH model within the DMFT combined with Wilson's
numerical renormalization group (NRG) \cite{bulla2}. The NRG
technique \cite{krishna1,wilson1} is particularly powerful in
that it is non-perturbative in nature so that it can cover the
whole parameter space and that it can probe the extremely small
energy scales like the narrow coherence peak and soft phonon mode
which emerge as the metal-insulator transitions are approached.

We will present a zero temperature phase diagram of unbroken
symmetry ground states with a focus on the nature of and
transitions between the ground states as $U$ and $g$ are varied.
The main results of the present work are: (i) The ground state is
a metallic (M) state when both $U$ and $g$ are small, but is a
bipolaron insulating (BPI) state when $g$ is large, and is a
Mott-Hubbard insulating (MHI) state when $U$ is large (see Fig.\
1). In the shaded region in Fig.\ 1, the insulating and metallic
solutions coexist, and the ground state is the one with a lower
energy. (ii) The phase transition between M and BPI occurs along
the dashed line that cuts through the coexistence region
connecting E and T of Fig.\ 1. The order parameter ($\big<n-1
\big> $, see below.) changes discontinuously at the transition
and the M-BPI transition is first order (Fig.\ 2). (iii) The
transition between M and MHI occurs along the outer boundary of
the coexistence region denoted by the solid line connecting T and
U2. It is second order as in the Hubbard model (Fig. 3). (iv) The
two kinds of insulators (BPI and MHI) are distinct and can $not$
be adiabatically connected. The transition between them is first
order, and occurs along $U=2g^2/\omega_0$ (Fig.\ 4. See below).
These points will be discussed in more detail below.

%{\it Hubbard-Holstein model.} --
The HH model is defined by
 \ba
 \label{hhm}
{\cal H} &=& -\frac{t}{\sqrt{q}} \sum_{<i,j>\sigma}
c_{i\sigma}^\dag
c_{j\sigma} +U \sum_i n_{i\ua} n_{i\da} \nonumber\\
&+&\omega_0 \sum_i a_i^\dag a_i + g \sum_i \left( a_i^\dag +a_i
\right) \left( n_i -1 \right),
 \ea
where $q$ is the nearest-neighbor coordination number,
$n_{i\sigma}=c_{i\sigma}^\dag c_{i\sigma}$ the electron density
operator for spin $\sigma$ at site $i$, and $n_i =\sum_\sigma
n_{i\sigma}$.
%The chemical potential $\mu=U/2$ corresponds to the
%half-filling we consider here.
The electrons have the onsite Coulomb repulsion $U$, and are
linearly coupled with the Einstein phonon of frequency $\omega_0$
with the onsite coupling constant $g$.
%The coupling is taken to be proportional to $(n_i -1)$,
%where the factor of $-1$ renders the problem particle-hole
%symmetric for the half-filled HH model.
The DMFT was employed to solve the HH model of Eq.\ (\ref{hhm}).
The DMFT maps a lattice model onto an effective single site
impurity model imbedded in a bath which is determined
self-consistently via iterations \cite{metzner,georges}. The
effective impurity problem was solved by the NRG technique in the
present study.

%Among the various methods applied to solve
%the effective impurity model, Wilson's NRG is particularly
%powerful in that (a) it is non-perturbative in nature so that it
%can cover the whole parameter space, (b) it can be applied to both
%zero and finite temperatures and to both the dynamical and static
%properties, and (c) it can resolve arbitrarily small energy scale
%around the Fermi energy \cite{bulla2,krishna1,wilson1}.

%{\it NRG+DMFT.} --
The NRG+DMFT has been successfully applied to
the half-filled Hubbard model to study the MIT at both $T=0$
\cite{bulla2} and finite temperatures \cite{bulla3}, and to the
half-filled Holstein model at $T=0$ \cite{meyer}. For the Hubbard
model, the ground state is insulating for $U> U_{c2}$ and
metallic for $U<U_{c1}$. Between $U_{c1}<U<U_{c2}$, the metallic
and insulating solutions coexist, and the metallic state has the
lower energy and is the ground state. The MIT occurs at $U_c =
U_{c2}$, and is 2nd order. For the Holstein model, the ground
state is insulating for $g> g_{c2}$ and metallic for $g<g_{c1}$.
Meyer {\it et al.} reported that $g_{c2}-g_{c1}$ is reduced as
the phonon frequency $\omega_0$ is decreased and $g_{c1}=g_{c2}$
for $\omega_0 = 0.05~ W$, where $W=4t$ is the bandwidth
\cite{meyer}. We generalized these lines of research and study
the HH model at half-filling with the NRG+DMFT incorporating the
improved method for calculating electron \cite{bulla-jpc98} and
phonon spectral function \cite{jeon}. We adopt the semi-circular
density of states which is realized in the Bethe lattice in
infinite dimensions.

\begin{figure}
\label{fig1}
\includegraphics[scale=0.35]{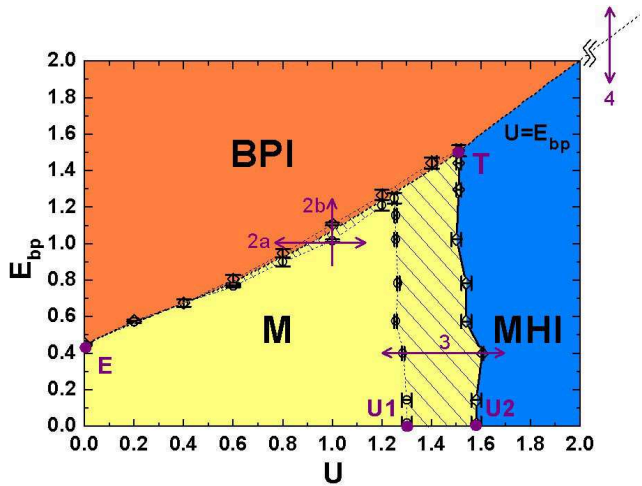}
%\includegraphics[scale=0.35]{fig1.jpg}
%\epsfig{file=fig1.eps,height=3cm,width=3cm,clip=,bbllx=0,bblly=0,bburx=3,bbury=3}
%\epsfig{scale=0.6,file=fig1.eps}
%\centerline{\epsfysize=2cm\epsfbox{fig1.eps}}
\caption{The zero-temperature metal-insulator transition phase
diagram of the HH model at half-filling. The cuts labeled with 2
(a \& b), 3, and 4 correspond, respectively, to the Figs.\ 2, 3,
and 4. The cuts 2a, 2b, 3, and 4 represent, respectively, the
calculations along the fixed $g=0.16$, $U=1.0$, $g=0.1$, and
$U=2.5$. The shaded area is the coexistence region where both the
metallic and insulating solutions exist. The dashed lines between
M and BPI, and between BPI and MHI states represent 1st order
phase transitions, and the solid line between M and MHI
represents 2nd order. See the text for details.}
\end{figure}

Our central result is the Fig.\ 1, where we present a zero
temperature phase diagram in the $U-E_{bp}$ plane of unbroken
symmetry ground states of the HH model at half-filling obtained
from the NRG+DMFT calculations. $E_{bp}=2g^2/\omega_0$ is the
bipolaron binding energy which is twice the polaron energy,
$E_{bp}=2E_p$. We take the bandwidth $W$ as the unit of energy
($W=1$).
%The detailed shape of the phase diagram depends on the
%remaining parameter $\omega_0$ of the HH model.
We take $\omega_0
= 0.05$ as with Meyer {\it et al.} Three distinct states are
found in the $U-E_{bp}$ plane: metal, bipolaron insulator, and
Mott-Hubbard insulator states. We will look at them in more
detail below. Specifically, we will address (i) the nature of M,
BPI, and MHI states, (ii) transition between M and BPI, (iii)
transition between M and MHI, and (iv) transition between the BPI
and MHI. The results obtained from the NRG+DMFT regarding these
points were summarized in the introduction above.

{\it (i) Nature of M, BPI, and MHI states.} -- The ground state is
a metallic state when both $U$ and $g$ are small. The M state
becomes more correlated as $g$ or $U$ is increased as reflected in
the decreasing quasiparticle weight $ z = \left[ 1-\partial
\Sigma_1(\omega)/\partial\omega \right]^{-1}_{\omega\ra 0},$ where
the subscript 1 (2) refers to the real (imaginary) part. $z$
vanishes at the outer boundary of the coexistence region.
%The M-BPI phase transition occurs before $z=0$ and is 1st
%order; M-MHI at $z=0$ and is 2nd order (see below.)
The BPI state is insulating because the attractive interaction
between the electrons mediated by the phonons binds two electrons
into a bipolaron as Capone and Ciuchi reported \cite{capone1}. The
MHI state, on the other hand, is insulating because of the
repulsion $U$ between two electrons at the same site.

An important point for understanding the overall feature of the
phase diagram is that the effective interaction between two
electrons in the HH model is given, after integrating the phonons
out, by
 \ba
V_{eff}(\omega) = U+\frac{2g^2 \omega_0}{\omega^2-\omega_0^2}.
 \label{eff}
 \ea
The interactions between electrons due to $g$ and $U$ compete each
other for $\omega \lesssim \omega_0$, which yields
$V_{eff}(\omega=0)=0$ along $U=E_{bp}$. The $U=E_{bp}$ line plays
a special role in the phase diagram in that all three states meet
at the triple point T lying on this line and the BPI-MHI
transition occurs along it, as we will discuss below. Also, the
critical $g_c$, which separates M and BPI along E and T, should
increase as $U$ is increased because the attractive interaction
due to $g$ has to overcome the repulsive interaction of $U$ to
bind two electrons into a bipolaron. This expectation is indeed
borne out as shown in Fig.\ 1. On the other hand, the $U_c$, which
is the solid line connecting T and U2, depends weakly on $g$.

\begin{figure}
\label{fig2}
\includegraphics[scale=0.30]{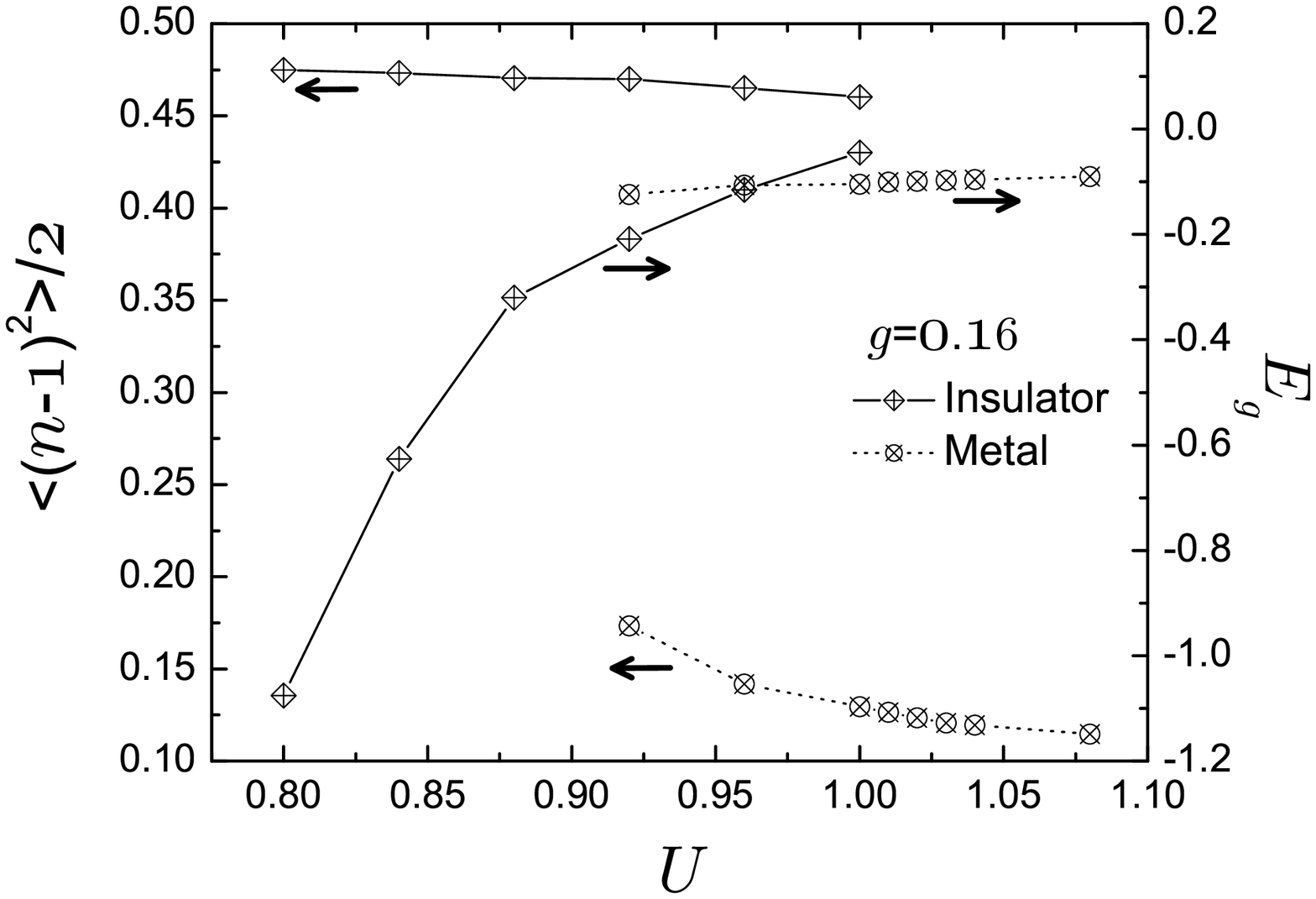}
\includegraphics[scale=0.30]{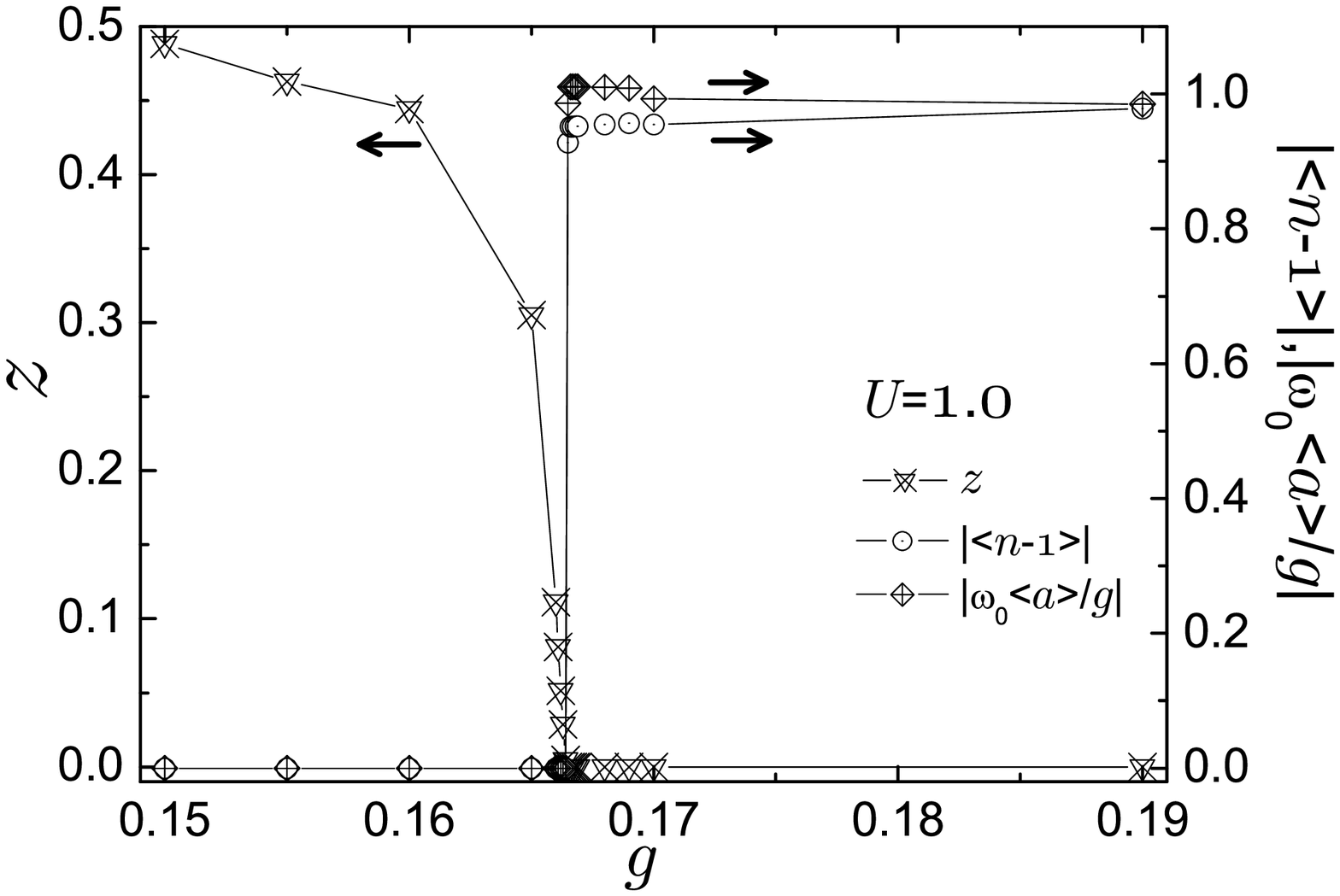}
\includegraphics[scale=0.33]{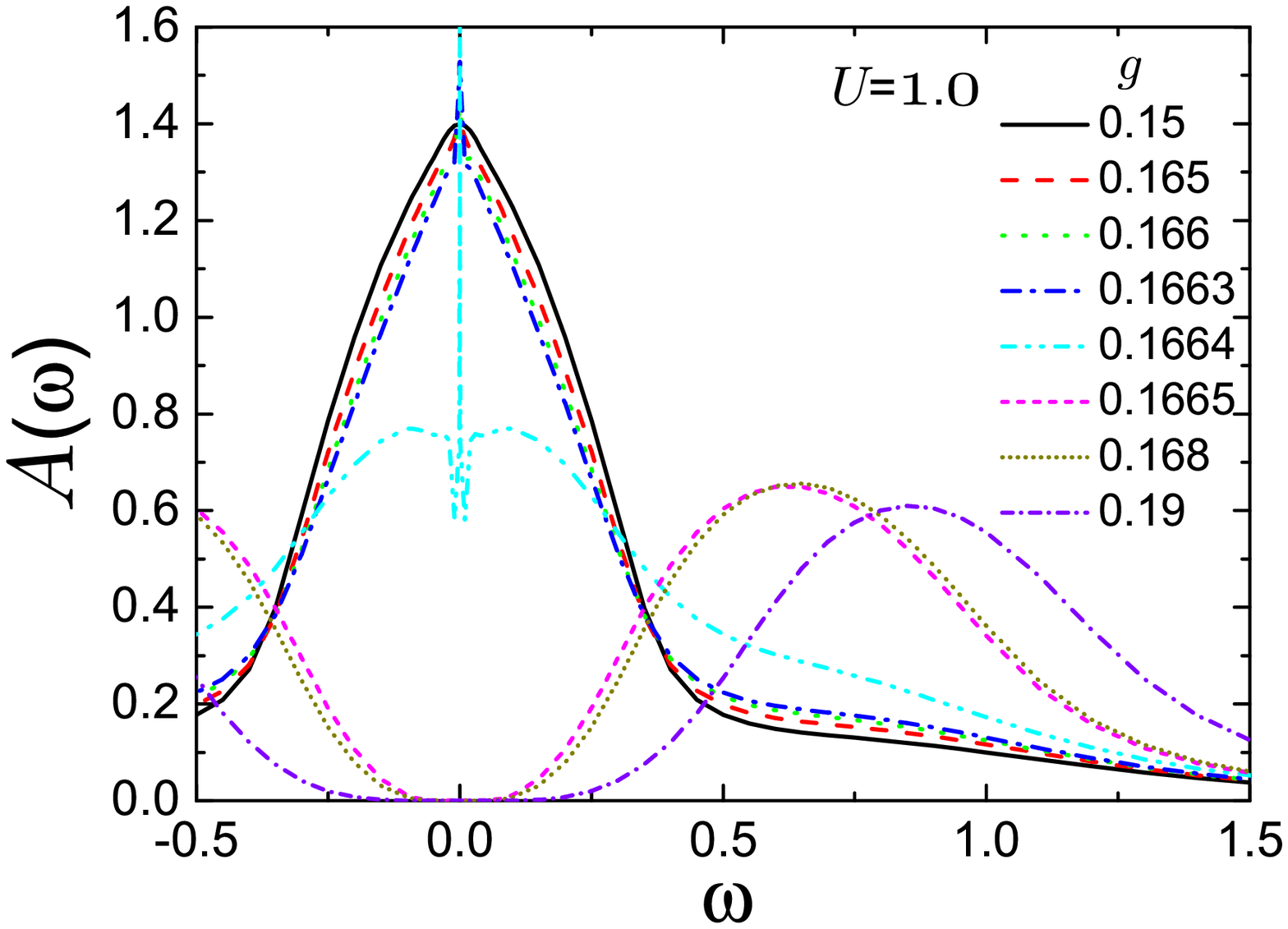}
\includegraphics[scale=0.33]{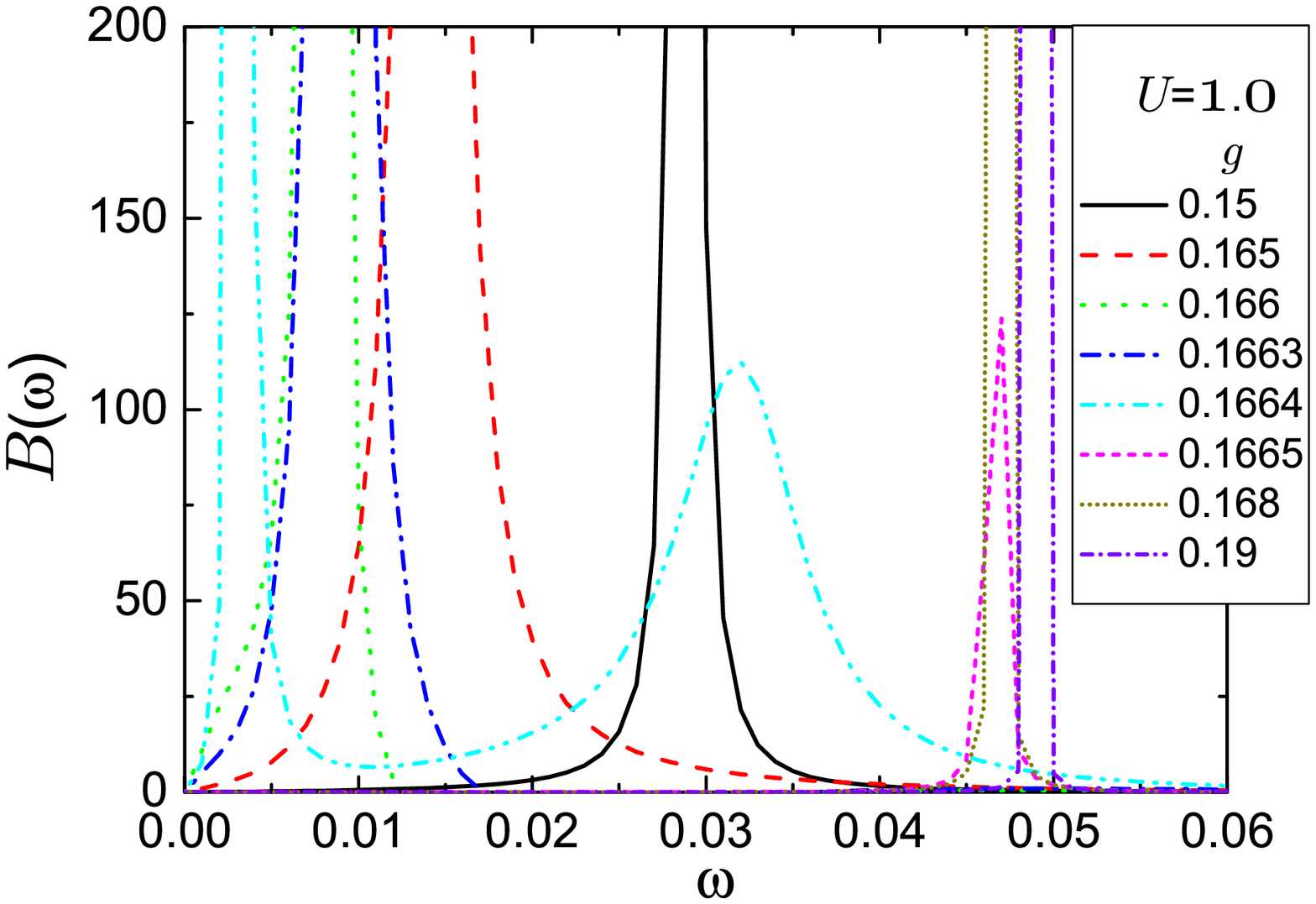}
\caption{(a) The $\big<(n-1)^2\big>/2$, which is equal to the
double occupancy $\big< n_{\ua} n_{\da}\big>$ in M state, and
ground state energy $E_g$ are plotted along the cut 2a with a
fixed $g=0.16$ in Fig.\ 1 as $U$ is increased (for M) or
decreased (for BPI). It can be read off from the plots that the
coexistence region is $0.925<U<1.0$, and the phase transition
occurs around $U=0.96$ where the energies of the M and BPI states
become the same. (b) Plots of $z$, $C$, and $\frac{\omega_0}{g}
\big<a\big>$ as $g$ is increased along the cut 2b of Fig.\ 1. In
Figs.\ (c) and (d), the electron and phonon spectral functions,
$A(\omega)$ and $B(\omega)$, respectively, are shown along the
same cut with (b). }
\end{figure}

%In the shaded area of Fig.\ 1, both the
%metallic and insulating solutions exist. The ground state is then
%the one with a lower energy.
%which follows
%from the following observations: (1) $dE_g /dU = \big < n_\ua
%n_\da \big>$, where $E_g$ is the ground state energy, (2) $\big<
%n_\ua n_\da \big>_{MHI} < \big < n_\ua n_\da \big>_{M}$,
%(3) $\big < n_\ua n_\da \big>_M$ and $\big< n_\ua n_\da
%\big>_{MHI}$ meet as $U\ra U_{c2}$, and the metallic and
%insulating solutions become identical. These observations also
%mean that the MIT at $U_{c2}$ is 2nd order \cite{moeller}.

For the coexistence region with $U>E_{bp}$, the arguments which
led to the conclusion of the MIT of the Hubbard model being a 2nd
order at $U=U_{c2}$ \cite{moeller} still apply and the M is the
ground state.
%The phase transition between M and MHI occurs along the outer
%boundary of the coexistence region denoted by the solid line
%connecting T and U2, and is 2nd order.
On the other hand, in the coexistence region with $E_{bp}>U$, the
$\big< n_\ua n_\da \big>_{BPI}$ and $ \big < n_\ua n_\da
\big>_{M}$, where $\big< \cal O\big>_{\it M (I)} $ stands for an
expectation value of an operator $\cal O$ in the metallic
(insulating) state, do not meet as shown in Fig.\ 2(a). This
implies that the M-BPI transition is 1st order. The energy
calculations in Fig.\ 2(a) show that the M and BPI solutions
cross each other inside the coexistence region.
%The phase transition between M and BPI occurs along the dashed line
%connecting E and T that cuts through inside the coexistence
%region.

{\it (ii) Transition between M and BPI.} -- The MIT of half-filled
Holstein model in infinite dimensions was discussed previously by
Meyer {\it et al.} \cite{meyer} and by Benedetti and Zeyher
\cite{benedetti}. The BPI state is insulating because electrons
are bound in bipolarons. Because we set the chemical potential
$\mu=U/2$, which prefers the electron density $\big<n\big>=1$, the
bipolaron formation may be accommodated by reconstructuring the
system into a phase separated state or a charge ordered state
where the doubly occupied and empty sites alternate in the real
space. This possibility is not allowed in the present DMFT work
because the unit cell consists of a single site. Instead, the
bipolaron instability shows up as degenerate ground states: In
the NRG+DMFT calculations, the ground state is 8 fold degenerate
in the BPI regime, with a one-to-one correspondence between the
two sets of $\big<n\big>-1 \approx 1$ and $\big<n\big>-1 \approx
-1$, which on averaging satisfies $\big<n\big>=1$. The order
parameters for the M-BPI may be taken as
 \ba
 \label{order}
C=\big<n-1\big>.
 \ea

Now, consider the cut 2b in Fig.\ 1 along which we increase $g$
with $U=1.0$. Let $g_{c1}$ and $g_{c2}$ be, respectively, the
lower and upper boundaries of the shaded coexistence region. As
$g\ra g_{c2}$, the quasiparticle weight $z$ and the renormalized
phonon frequency $\Omega$ approach 0 continuously, but, the order
parameter $C$ increases from 0 to a finite value discontinuously,
as shown in Fig.\ 2(b) and (d). The softening phonon mode is a
manifestation of a lattice instability just like structural phase
transitions. A stability is restored by a condensation of the
unstable mode. It results in a non-zero expectation value of the
phonon operator ($\big<a\big>\ne 0$), which may be taken as an
order parameter for the M-BPI transition. The phonon hardens back
to the bare mode as $g$ is increased above $g_{c2}$ because the
screening is not effective in an insulating state. The $a$ and
$(n-1)$ are linearly coupled in the Hamiltonian of Eq.\
(\ref{hhm}), and a zero/nonzero $\big<a\big>$ is expected to lead
to a zero/nonzero $C$. Therefore, an equally good order parameter
is the $C$ of Eq.\ (\ref{order}).

In Fig.\ 2(b), we show $z$, $C$, and $\frac{\omega_0}{g}
\big<a\big>$ with increasing $g$. Note that $\big<n-1 \big>
\approx\frac{\omega_0}{g}\big<a\big>$ as expected from a simple
mean-field theory. In Fig.\ 2(c), we plot the renormalized
electron spectral function, $A(\omega)=-\frac1\pi Im \llangle c,
c^\dag \rrangle_\omega$, to show the MIT as $g$ is increased
along the cut 2b. In Fig.\ 2(d), we plot the renormalized phonon
spectral function, $B(\omega)=-\frac1\pi Im \llangle a+a^\dag,
a+a^\dag \rrangle_\omega$, calculated as $g$ is increased using
the method in \cite{jeon}. Note that at $g=0.1664$ which is close
to $g_{c2}=0.1665$, the phonon mode splits into two components as
reported by Jeon $et~ al.$ for the Anderson-Holstein model: one
component develops into the soft mode and the other hardens back
to the bare mode \cite{jeon}. But, in the coexistence region
where $g_{c1}$ and $g_{c2}$ are not close, the 1st order MIT from
M to BPI preempts the emergence of the soft mode as $g$ is
increased because the MIT occurs at $g_c$ ($g_{c1}<g_c<g_{c2}$).

The M-BPI transition may be understood from the Ginzburg-Landau
theory. The free energy may be written in terms of the the local
phonon coordinate $\xi = \sqrt{\frac{\hbar}{2M\omega_0}} (a^\dag
+a)$ as
 \ba
F = \frac12 \alpha\xi^2 +\frac14 \beta\xi^4 +\frac16 \gamma\xi^6,
 \ea
with $\gamma>0$. The phase transition can be 1st order for
$\beta<0$. In the context of the M-BPI transition, the
$g_{c1}<g<g_{c2}$ regime corresponds to
$\beta^2/4\gamma>\alpha>0$, for which three local minima exit as a
function of $\xi$. The $g_{c1}$, $g_c$, and $g_{c2}$ correspond,
respectively, to $\alpha=\beta^2/4\gamma$, $3\beta^2/16\gamma$,
and 0. Although the $\alpha$ and $\beta$ are complicated functions
of $U$ and $g$, the DMFT phase diagram of Fig.\ 1 indicates that
$\alpha$ and $\beta$ are mainly determined by, respectively, $g$
and $U$. This observation suggests an interesting possibility that
the M-BPI transition may become 2nd order for small values of $U$
where $\beta \geq 0$, and a quantum tricritical point exists
between E and T. This possibility will be explored further in a
subsequent study.

\begin{figure}
\label{fig3}
\includegraphics[scale=0.30]{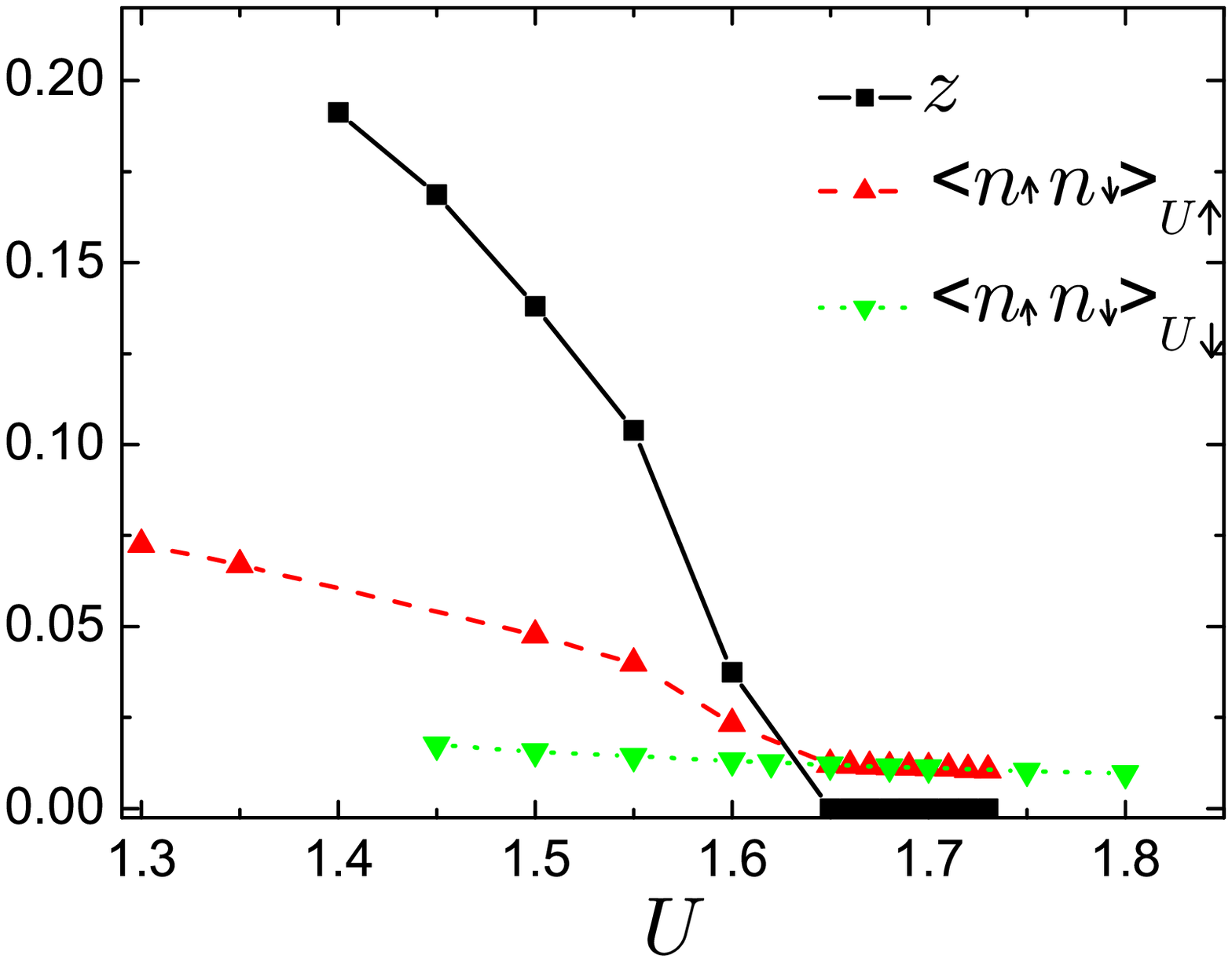}
\includegraphics[scale=0.33]{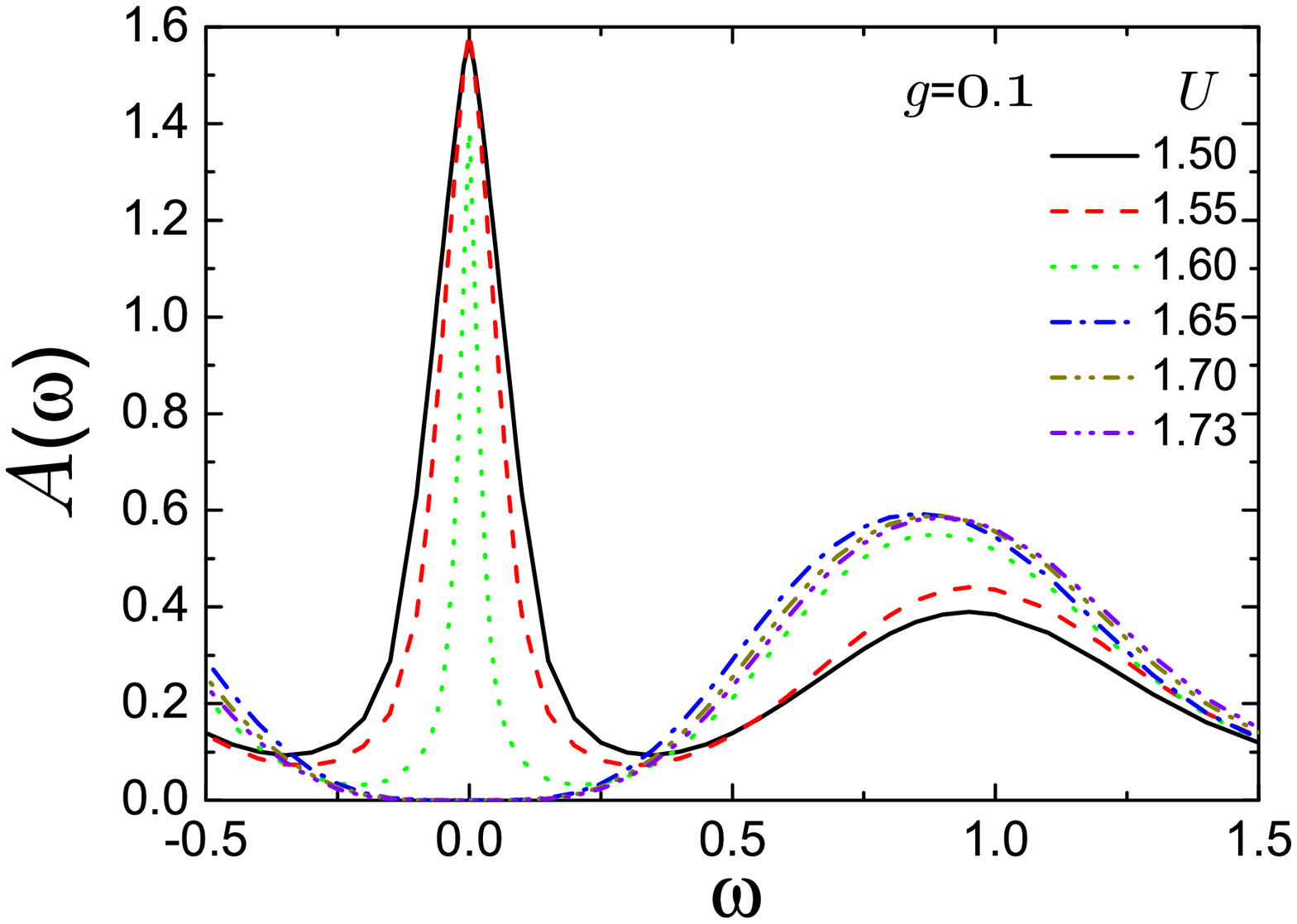}
\caption{In Fig.\ (a), we plot $z$ as $U$ is increased, and
$\big< n_\ua n_\da \big>$ as $U$ is increased and decreased along
the cut 3 of Fig.\ 1. In Fig.\ (b), the electron spectral
functions $A(\omega)$ are shown as $U$ is increased along the
same cut. }
\end{figure}

{\it (iii) Transition between M and MHI.} --
%As the phase
%transition line connecting T and U2 is approached, an MIT occurs.
%The $z$ approaches 0 continuously as $U\ra U_{c2}$ from below.
As the MIT from M to BPI is driven by growing charge
susceptibility, the one from M to MHI is driven by growing spin
susceptibility.
%To characterize the M-MHI transition, we
%calculate $I_2$, the intensity of the delta function in the
%imaginary part of the self-energy, $\Sigma_2$. For insulators,
%$\Sigma_2 (\omega) = -\pi I_2 \delta(\omega)$ within a gap
%\cite{note2}. The $I_2$ is a generic order parameter for
%insulators \cite{georges}.
In Fig.\ 3(a), we plot $z$ as $U$ is increased, and $\big< n_\ua
n_\da \big>$ as $U$ is increased and decreased for a fixed
$g=0.1$. $z$ goes to 0 continuously, and $\big< n_\ua n_\da
\big>_M = \big< n_\ua n_\da \big>_{MHI}$ as $U\ra U_{c2}$. The
M-MHI transition at $U_c=U_{c2}$ is, therefore, 2nd order.
%It is
%interesting to note that $I_2$ is non-zero at $U=U_{c2}$, which
%implies that $I_2$ changes discontinuously at $U_c$ although the
%M-MHI transition is 2nd order according to the classification of
%the phase transition by the discontinuity of the
%$n$-th derivative of the free energy.
The phonons are hardly renormalized (not shown) because the charge
degree of freedom is not soft unlike the MIT between M and BPI. In
Fig.\ 3(b), $A(\omega)$ is plotted to show the MIT as $U$ is
increased.

The $U_c$ does depend on $g$, although weakly, as can be seen from
the Fig.\ 1, which implies that the M-MHI transition is coupled
with the phonons. This raises an interesting possibility that some
discrepancies between the experimental observations on V$_2$O$_3$
systems and DMFT results of the Hubbard model may be resolved in
terms of the electron-phonon coupling \cite{limelette}.

{\it (iv) Transition between BPI and MHI.} -- Now, we turn to an
interesting question about the two kinds of insulators: Are BPI
and MHI different, or, put differently, can the two insulating
states be adiabatically connected? Along the line $U=E_{bp}$, the
effective interaction in the static limit vanishes. One might then
naively expect that a metallic state is the ground state along
this line all the way to $U=E_{bp} \ra \infty$. One can show,
however, that for $U=E_{bp} \ra \infty$ the ground state is an
insulating state using the exact atomic limit Green's function
\cite{jeon,hewson}, as found from the detailed NGR+DMFT
calculations.

In Fig.\ 4, we show $C$ as $g$ is varied for $U=2.5$. The $C$
changes discontinuously exactly at $E_{bp}=U$ and does not exhibit
any other non-analytic behavior as $g$ is varied. We checked
another point on the line $E_{bp}=U$ by varying $U$ with a fixed
$g=0.3$. The transition occurs exactly at $E_{bp}=U$ (not shown).
This confirms that the two insulating states are distinct, and the
phase transition between them occurs along the line of $U=E_{bp}$
and is 1st order. The point T in Fig.\ 1 is a triple point where
two 1st order and one 2nd order phase transition lines meet.

\begin{figure}
\label{fig4}
\includegraphics[scale=0.30]{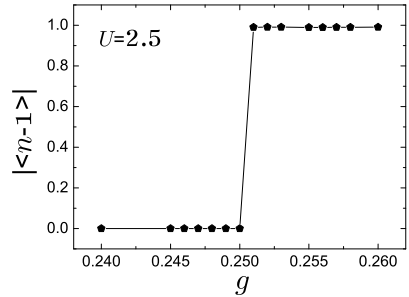}
\caption{The order parameters $C$ is plotted as a function of $g$
for the fixed $U=2.5$ corresponding to the cut 4 of Fig.\ 1. The
phase transition between BPI and MHI occurs along $U=E_{bp}$.}
\end{figure}

%{\it Summary and Outlook.} -- We presented in this paper a zero
%temperature metal-insulator transition phase diagram of the
%Hubbard-Holstein model at half-filling using NRG+DMFT. The nature
%of the metallic, bipolaron insulating, and Mott-Hubbard insulating
%states, and the transitions among them were discussed. We believe
%this is a major step towards systematic understanding of the
%intriguing interplay between the electron-electron and
%electron-phonon interactions.
%
%The
%parameters of the model are $t$, $U$, $\omega_0$, and $g$. Other
%control parameters will be the chemical potential (or the
%electron density) and the temperature. We presented the phase
%diagram in the $U-E_{bp}$ plane with a fixed $\omega_0/t=0.2$.
%
%The remaining interesting questions will be how the phase diagram
%and nature of transitions will change as $\omega_0$, the electron
%density, or the temperature is changed. Also interesting will be
%to allow the symmetry broken states such as the
%antiferromagnetic, charge ordering, and superconducting states.
%Works in these directions are in progress and will be reported in
%separate publications.

We would like to thank Tae-Suk Kim, Yunkyu Bang, Ralf Bulla, Alex
Hewson, and Dietrich Meyer for helpful comments and discussions.
This work was supported by the Korea Science \& Engineering
Foundation (KOSEF) through grant No.\ R01-1999-000-00031-0 and
Center for Nanotubes and Nanoscale Composites (CNNC), and by the
Ministry of Education through BK21 SNU-SKKU Physics program. JHH
and HCL acknowledge the support from Center for Strongly
Correlated Materials Research (CSCMR).


\begin{thebibliography}{99}

%\bibitem{millis1} A. Deppeler and A. J. Millis, Phys. Rev. {\bf
%65}, 100301 (2002).

\bibitem{imada} M. Imada, A. Fujimori, Y. Tokura, Rev. Mod.
Phys. {\bf 70}, 1039 (1998).

\bibitem{metzner} W. Metzner, D. Vollhardt, Phys. Rev. Lett. {\bf 62}, 324
(1989); $ibid$, 1066 (1989).

\bibitem{georges} A. Georges, G. Kotliar, W. Krauth, M. J.
Rozenberg, Rev. Mod. Phys. {\bf 68}, 13 (1996).

\bibitem{bulla2} R. Bulla, Phys. Rev. Lett. {\bf 83}, 136 (1999).

\bibitem{krishna1} H. R. Krishna-murthy, J. W. Wilkins, K. G.
Wilson, Phys. Rev. B {\bf 21}, 1003 (1980).

%\bibitem{kristina2} H. R. Krishna-murthy, J. W. Wilkins, and K. G. Wilson,
%Phys. Rev. B {\bf 21}, 1044 (1980).

\bibitem{wilson1} K. G. Wilson, Rev. Mod. Phys. {\bf 47}, 773
(1975).

%\bibitem{bulla1} R. Bulla, Adv. Solid State Phys. {\bf 40}, 169 (2000).

\bibitem{bulla3} R. Bulla, T. A. Costi, D. Vollhardt, Phys. Rev. B {\bf 64}, 045103
(2001).

%\bibitem{mahan} G. D. Mahan, {\it Many-Particle Physics}, 3rd ed.
%(Plenum, New York, 2000).

\bibitem{meyer} D. Meyer, A. C. Hewson, R. Bulla,
Phys. Rev. Lett. {\bf 89}, 196401 (2002).

\bibitem{bulla-jpc98} R. Bulla, A. C. Hewson, Th. Pruschke, J. Phys.:
Condens. Matter {\bf 10}, 8365 (1998).

\bibitem{jeon} G. S. Jeon, T. H. Park, H. Y. Choi,
Phys. Rev. B {\bf 68}, 045106 (2003).

\bibitem{capone1} M. Capone, S. Ciuchi, Phys. Rev. Lett. {\bf 91}, 186405 (2003).

\bibitem{moeller} G. Moeller, Q. Si, G. Kotliar, M. Rozenberg,
D. S. Fisher, Phys. Rev. Lett. {\bf 74}, 2082 (1995).

\bibitem{benedetti} P. Benedetti, R. Zeyher, Phys. Rev. B {\bf
58}, 14320 (1998).

%\bibitem{note2} This may be understood as follows: The
%frequency-dependent conductivity can be written
%phenomenologically as $ \sigma_1 (\omega) = \frac{ne^2}{m}
%\frac{\tau^{-1}(\omega)}{\omega^2+\tau^{-2}(\omega)},$ where
%$\tau^{-1}(\omega)\approx -2\Sigma_2(\omega)$. In order that
%$\sigma_1(\omega)=0$ within the gap, $\tau^{-1}\ra\infty$ at
%$\omega=0$ and $\tau^{-1}=0$ for $\omega\ne 0$, which means that
%$\Sigma_2$ should have a delta function in an insulating state at
%$T=0$.

\bibitem{limelette} P. Limelette {\it et al.}, Science {\bf 302}, 89 (2003).

\bibitem{hewson} A. C. Hewson, D. Meyer, J. Phys.: Condens.
Matter {\bf 14}, 427 (2002).

%\bibitem{note1} We obtain
%$g_{c2}(U=0)-g_{c1}(U=0) \approx 0.002$ for $\omega_0 = 0.05$
%using the NRG discretization parameter $\Lambda=2.0$. It will be
%of interest and importance to check if we can have a finite region
%in the $U-g$ plane where $g_{c1}=g_{c2}$ by decreasing $\omega_0$
%or taking $\Lambda\ra 1$ limit.


%\bibitem{Lea01}
%A.~Lanzara, P.~V. Bogdanov, X.~J. Zhou, S.~A. Keller, D.~L. Feng,
%E.~D. Lu, T.~Yoshida, H.~Eisaki, A.~Fujimori, K.~Kishio, J.-I.
%Shimoyama, T.~Nodaand~S. Uchida, Z.~Hussain, and Z.-X. Shen,
%Nature {\bf 412}, 510 (2001).

%\bibitem{Ram97}
%A.~P. Ramirez, J. Phys.: Condens. Matter {\bf 9}, 8171 (1997).

%\bibitem{MLS95}
%A.~J. Millis, P.~B. Littlewood, and B.~I. Shraiman, Phys. Rev.
%Lett. {\bf 74}, 5144 (1995).


\end{thebibliography}
\end{document}